# Suppression of magnetoresistance in thin WTe$_2$ flakes by surface oxidation

J. M. Woods[1,2,‡], J. Shen[1,2,+,‡], P. Kumaravadivel[1,2], Y. Pang[3], Y. Xie[1,2,4], G. A. Pan[5], M. Li[6], E. I. Altman[4,7], L. Lu[3], J. J. Cha[1,2,4,*]

[1] Department of Mechanical Engineering and Materials Science, Yale University, New Haven, CT 06511, USA

[2] Energy Sciences Institute, Yale West Campus, West Haven, CT 06516, USA

[3] Institute of Physics, Chinese Academy of Sciences, Beijing, China

[4] Center for Research on Interface Structures and Phenomena, Yale University, New Haven, CT 06511, USA

[5] Department of Physics, Yale University, New Haven, CT 06511, USA

[6] Materials Characterization Core, Yale West Campus, West Haven, CT 06516, USA

[7] Department of Chemical & Environmental Engineering, Yale University, New Haven, CT 06520, USA

[+] present address: J.Shen-1@tudelft.nl

‡These authors contributed equally to this work.

CORRESPONDING AUTHOR FOOTNOTE:

*Corresponding author. E-mail: judy.cha@yale.edu

Telephone number: +1 (203) 737-7293, Fax number: (203) 432-6775

Abstract:

Recent renewed interest in layered transition metal dichalcogenides stems from the exotic electronic phases predicted and observed in the single- and few-layer limit. Realizing these electronic phases requires preserving the desired transport properties down to a monolayer, which is challenging. Here, using semimetallic $WTe_2$ that exhibits large magnetoresistance, we show that surface oxidation and Fermi level pinning degrade the transport properties of thin $WTe_2$ flakes significantly. With decreasing $WTe_2$ flake thickness, we observe a dramatic suppression of the large magnetoresistance. This is explained by fitting a two-band model to the transport data, which shows that mobility of the electron and hole carriers decreases significantly for thin flakes. The microscopic origin of this mobility decrease is attributed to a ~ 2 nm-thick amorphous surface oxide layer that introduces disorder. The oxide layer also shifts the Fermi level by ~ 300 meV at the $WTe_2$ surface. However, band bending due to this Fermi level shift is not the dominant cause for the suppression of magnetoresistance as the electron and hole carrier densities are balanced down to ~ 13 nm based on the two-band model. Our study highlights the critical need to investigate often unanticipated and sometimes unavoidable extrinsic surface effects on the transport properties of layered dichalcogenides and other 2D materials.

Keywords:

2D Layered Materials, Large Magnetoresistance, $WTe_2$, Electronic Transport, Fermi Level Pinning

Main Text:

Exotic electronic phases found in layered transition metal dichalcogenides (TMDCs), such as charge density waves,[1-3] topological surface states,[4] and 2D superconductivity,[2,5] make TMDCs promising for a wide range of novel thin film electronics. In the single layer limit, additional electronic phases such as the quantum spin Hall state[6] are predicted, which can be harnessed for spintronic and low-dissipation electronic applications. Recent progress in the ability to build heterostructures by stacking chalcogenide single layers with different electronic phases offers exciting opportunities for realizing interface- and proximity-induced electronic phases not possible in bulk forms.[7] However, a critical requirement for device applications is to retain these predicted electronic phases and transport properties in the single layer limit. Often, this is not the case, as demonstrated by the presence of a 2D electron gas with low mobility at the surface of topological insulators such as $Bi_2Se_3$ and by surface oxidation of many 2D materials.[8-10] Encapsulating layered materials using boron nitride can be effective in preserving transport properties as shown for graphene,[11, 12] but is still limited to individual nanoscale devices. Systematic studies of thickness-dependent transport properties of TMDCs are crucial for implementing practical applications based on these materials.

Among the TMDCs $WTe_2$ is a candidate for a type II Weyl semimetal[13] and has recently received much attention largely due to its non-saturating large magnetoresistance (MR).[14] This large MR has been ascribed to the near perfect electron and hole charge compensation,[14] further supported by band structure calculations,[15] angle-resolved photoemission spectroscopy (ARPES) measurements,[16] and analysis of Shubnikov-de Hass oscillations.[17-19] A classical two-band model is sufficient to describe the observed large MR despite the fact that the actual band structures are more complicated with four Fermi pockets along the Γ-X axis in the 1$^{st}$ Brillouin zone.[19, 20] In addition, monolayer $WTe_2$ in the 1T' phase is predicted to exhibit the quantum spin Hall effect.[6] Further theoretical calculation of layered $WTe_2$ predicts that the extremely large MR, and therefore the near-perfect charge compensation and high mobility, should be preserved down to the monolayer limit.[15] The actual preservation of these electronic properties for fabricated devices of monolayer $WTe_2$ is vital in harnessing the exotic transport behavior[6].

In contrast to these theoretical predictions, here we show that the MR of thin, mechanically exfoliated $WTe_2$ flakes decreases significantly compared to that of a bulk crystal. This is in agreement with the recent report by L. Wang *et al*.,[21] which also showed dramatic suppression of the MR in WTe$_2$ thin flakes down to 6 layers. The suppression of the MR was attributed to decreased mobility and increased disorder due to surface degradation but a microscopic origin was not investigated. Here, we reveal the microscopic origin for MR suppression by correlating transport measurements with the microstructure of WTe$_2$ flakes from 145 nm down to 2.5 nm (~ 3 layers). We find that the microscopic source is a ~ 2 nm amorphous surface oxide layer. The work function of the oxide layer is found to be ~300 meV less than that of clean WTe$_2$, suggesting a potential charge imbalance between the electron and hole carriers of WTe$_2$ at the WTe$_2$ surface due to downward band bending by the Fermi level pinning, which would lead to suppression of MR. However, due to the limited spatial extent of the band bending, estimated to be ~ 1 nm by Poisson's equation, the Fermi level pinning is not the dominant cause behind the degradation of MR. A much larger factor is the increased disorder induced by the surface oxide layer, reflected in the transport data as a decrease in mobility and weak anti-localization.

Electronic transport measurements of WTe$_2$ thin flakes, mechanically exfoliated from bulk crystals (Supporting Information Figure S1-3), were taken from 300 K to 2 K with magnetic fields up to 9 T. Figure 1(a)-(e) show the temperature-dependent longitudinal resistance, $R_{xx}(T)$, with and without a 9 T magnetic field perpendicularly applied to the basal plane of WTe$_2$. In the absence of a magnetic field, $R_{xx}(T)$ shows the expected semi-metallic behavior with the residual resistance ratio ($RRR$) defined as $R_{xx}(300K)/R_{xx}(1.8K)$. When field-cooled at 9 T, $R_{xx}$ increases at low temperature due to the emergence of the large MR, with $T_{on}$ defined as the temperature at which $R_{xx}$ starts to increase. Both $T_{on}$ and

*RRR* start to decrease significantly for flakes thinner than ~ 40 nm (Figure 1(f)). Figure 1(g)-(k) show the corresponding MR curves measured at 1.8 K with MR defined as *[R(B)-R(0)]/R(0)*. With decreasing WTe$_2$ thickness, MR decreases linearly (Figure 1(l)). These observations clearly indicate that the transport properties of WTe$_2$ degrade with decreasing WTe$_2$ thickness.

Strong correlations between the crystal quality and MR[22] as well as the carrier doping and the MR[23] have been shown using WTe$_2$ bulk crystals. Thus, the observed MR suppression suggests either degradation of crystal quality or changes in transport property such as the carrier density and mobility for thin WTe$_2$ flakes. We measure the Hall resistance *R$_{xy}$(B)* of the flakes from 145 nm down to 2.5 nm to find the origin of the MR suppression (Figure 2 and Figure S4). For all thicknesses, strong non-linearity is observed in *R$_{xy}$*, suggesting that at least two carrier types contribute to the Hall curves. In WTe$_2$ the Fermi level cuts through the conduction and valence bands,[14] so both electron and hole carriers are expected to contribute to transport. Therefore, we use the classical two-band model for resistivity ($\rho_{xx}$ and $\rho_{xy}$ in equation (1)[24]) to fit the *R$_{xx}$* and *R$_{xy}$* curves, with the appropriate geometrical factors considered, assigning one band to electrons and the other to holes. Supporting Information Figure S5 shows the fitted curves, which overlap well with the measured data down to the 13 nm flake.

$$\begin{cases} \rho_{xy} = \dfrac{(n_h\mu_h^2 - n_e\mu_e^2)B + \mu_h^2\mu_e^2(n_h - n_e)B^3}{e[(n_h\mu_h + n_e\mu_e)^2 + (n_h - n_e)^2\mu_h^2\mu_e^2 B^2]} \\ \rho_{xx} = \dfrac{(n_h\mu_h + n_e\mu_e) + (n_h\mu_h\mu_e^2 + n_e\mu_e\mu_h^2)B^2}{e[(n_h\mu_h + n_e\mu_e)^2 + (n_h - n_e)^2\mu_h^2\mu_e^2 B^2]} \end{cases} \quad (1)$$

Figure 2(f) shows the electron and hole carrier densities, $n_e$ and $n_h$, obtained from the fit as a function of WTe$_2$ thickness. Figure 2(g) shows the corresponding mobility values, $\mu_e$ and $\mu_h$. Figure 2(h) shows the charge carrier density ratio, $n_e/n_h$. We see that as thickness decreases the mobility and the total charge concentration decrease, but the ratio of the electron to hole densities remains close to one down to 13 nm. The two-band model does not describe well the transport data of the 6.3 nm and 2.5 nm flakes because of the weak anti-localization that starts to dominate in these thin flakes (Figure 1(k) and Figure S5(j)). Temperature-dependent MR and Hall studies for 115 nm and 13 nm WTe$_2$ flakes are shown in Figure S6. Based on the fit results using the two-band model, we attribute the suppression of MR with decreasing WTe$_2$ flake thickness (Figure 1(l)) to the decrease in electron and hole mobility values, and not to the charge imbalance as the electron and hole carrier densities are nearly equal down to the 13 nm WTe$_2$ flake.

The dramatic decrease in the mobility values suggests presence of a 2D impurity band with very low mobility that induces strong scatter in thin flakes. The effect of the 2D impurity band to the transport properties of WTe$_2$ thin flakes is studied by analyzing Shubnikov-de Hass (SdH) oscillations as a function of WTe$_2$ flake thickness. Figure 3(a) shows $R_{xx}(B)$ plotted in B$^2$, which clearly shows the quadratic dependence of MR on B, as expected from Equation (1) for $\rho_{xx}$ when $n_h \approx n_e$, and also shows the suppression of MR with decreasing thickness. From the second derivative of $R_{xx}(B)$, $d^2R_{xx}/dB^2$ (Figure 3(b)), SdH oscillations are observed down to the 13 nm thick sample, indicating the presence of high mobility carriers. More importantly, the field $B_Q$ at which the SdH oscillations start to emerge remains constant at ~ 4.7 T for all thicknesses down to 13 nm. Using the condition $\mu_{SdH} \cdot B_Q \sim 1$ we obtain a constant mobility for the high mobility carriers. This is in contrast to the decreasing electron and hole mobilities obtained from the Hall curves (Figure 2(g)). The average mobility obtained from the Lorentz law, $MR \approx (\mu_{avg}B)^2$, is plotted together with $\mu_{SdH}$, for comparison, as a function of WTe$_2$ thickness (Figure 3(c)). The systematic decrease in $\mu_{avg}$ suggests increased scattering in thinner WTe$_2$ flakes due to the 2D impurity band near the surface. The constant $\mu_{SdH}$ suggests that the 2D impurity band does not scatter the high mobility carriers down to 13 nm. The role of the 2D impurity band on $\mu_{avg}$ is further substantiated from the change in conductivity (and corresponding resistivity) with decreasing thickness (Figure 3(d)).

To reveal the microscopic origin of the 2D impurity band, we examine a WTe$_2$ flake by cross-sectional transmission electron microscopy (TEM), energy dispersive X-ray spectroscopy (EDX), and X-ray photoelectron spectroscopy (XPS). Figure 4(a) shows the cross-sectional TEM image of a WTe$_2$ flake transferred onto a SiO$_2$/Si substrate, in which WTe$_2$ layers are clearly resolved (dotted red box). Below the crystalline WTe$_2$, we observe two different amorphous layers with a distinct contrast in intensity. The bottom layer of lighter contrast corresponds to the SiO$_2$ substrate, confirmed by EDX. Chemical analysis of the darker amorphous layer (dotted green box) indicates the presence of oxygen, tellurium, and tungsten (Figure 4(b)), whereas the oxygen peak is absent in the crystalline WTe$_2$ region. Thus, we conclude that a ~ 2 nm thick oxide layer forms on WTe$_2$ flakes. Oxide formation can be detected within a few hours by comparing XPS spectra of WTe$_2$ flakes, freshly cleaved/cleaned and after being left in air for 3.5 hours (Figure 4(c,d)). The clean WTe$_2$ flakes were obtained by sputtering the surface of freshly cleaved WTe$_2$ flakes for ~2 min with Ar at 5 keV, which completely removed the oxide layer. Pronounced W-$O_x$ and Te-$O_x$ peaks are observed after 3.5 hours in air, which were not present in the freshly cleaved and sputtered sample. The XPS data corroborates the observation that the amorphous layer observed in the cross-sectional TEM image is indeed a surface oxide of $WTe_2$.

The surface oxide layer induces disorder that manifests itself as a 2D impurity band with low mobility, leading to the suppression of MR for thinner flakes. In addition, the surface oxide layer can induce downward band bending near the surface of $WTe_2$ by pinning the Fermi level. This was measured by ultraviolet photoelectron spectroscopy (UPS). Figure 4(e) shows the secondary edge of clean (black) and air-exposed (red) $WTe_2$ flakes. After exposure to air, the secondary edge increases by about 300meV, which indicates a corresponding decrease in the work function from 4.5 eV to 4.2 eV. For the $WTe_2$ near the surface oxide, the bands are bend (Figure 4(f)). Using Poisson's equation and the dielectric constant of $WTe_2$[25] we determine the spatial extent of the band bending to be about 1 nm. This limited spatial extent of the band bending explains why the charge compensation holds down to a 13 nm $WTe_2$ flake. Full UPS profiles and details of Poisson's equation calculations are shown in Figure S7. Additionally, the ~ 2 nm thick oxide layer induces strong disorder for very thin $WTe_2$ flakes so that a localization effect should be discernible. This is indeed the case for the ~ 3 layer $WTe_2$ flake where a pronounced weak antilocalization (WAL) signal is observed due to the induced disorder (Figure 1(k)). Moreover, WAL is expected due to the strong spin-orbit coupling in $WTe_2$. We fit the two-dimensional Hikami-Larkin-Nagaoka localization model[26] to the observed WAL and obtain a $B_\phi$ of 0.074T for fixed $B_{SO}$ above 3.5 T (Figure S8).

Transport degradation due to surface oxidation is a major issue that affects most of the transition metal dichalcogenide family and other 2D materials. In the topological insulators $Bi_2Se_3$ and $Bi_2Te_3$, a disordered 2D gas at the surface complicates the analysis of the topological surface states.[8, 9] In black phosphorus, which possesses high mobility and anisotropic optical properties due to its crystal structure, fast surface oxidation significantly degrades device performance.[27, 28] Here, the observed degradation in transport properties of thin $WTe_2$ flakes is due to disorder induced by surface oxidation, leading to suppression of the large MR in $WTe_2$. In addition, our finding suggests that the predicted quantum spin Hall state may not be observed in unprotected single layer $WTe_2$ unless an effective encapsulation layer such as boron nitride is used.

## Methods

### WTe$_2$ synthesis

WTe$_2$ crystals were grown by a chemical vapor transport (CVT) method. 0.55 g of a WTe$_2$ source (American Elements, 99.999%) and 80 mg of an I$_2$ transport agent were placed in a quartz tube sealed at one end. The tube was purged with argon gas multiple times, and sealed with a base pressure below 50 mTorr. The sealed quartz vessel was placed in a two zone furnace. Over the course of 6 hours, the 'hot' end with the source powder was ramped up to 950 $^0$C and the 'cold' end was ramped up to 800 $^0$C. The furnace was held at these temperatures for 3 days. Upon cooling, WTe$_2$ crystals formed at the 'cold' end of the tube. These bulk crystals were then mechanically exfoliated on 285 nm SiO$_2$/Si substrates using the standard tape method to obtain thin WTe$_2$ flakes.

### Device fabrication and transport measurements

Devices were fabricated by standard e-beam lithography using a Vistec EBPG 5000+. 5/200 nm thick Cr/Au electrical contacts were deposited by thermal evaporation (MBraun MB-EcoVap). The low temperature and magnetic field transport measurements were performed soon after the lift-off process, using a Quantum Design Dynacool physical property measurement system equipped with a 9 T magnet and at a base temperature of 1.8 K. The thickness of the WTe$_2$ flakes was measured by a Bruker Fastscan AFM directly after the transport measurements to minimize the exposure of flakes to air.

**Structure characterizations**

The morphology and chemical composition were characterized by TEM/STEM (FEI Technai Osiris 200kV). Cross-sectional TEM samples were prepared by focused ion beam (Zeiss Crossbeam 540) at Carl Zeiss Microscopy, LLC. Planar samples were obtained by sonicating $WTe_2$ flakes in methanol and drop-casting onto a TEM grid.

XPS and UPS (He I illumination, E=21.22eV) measurements (PHI 5000 Versa Probe II) were conducted to determine the chemical and electrical nature of $WTe_2$ before and after exposure to air. For clean $WTe_2$ flakes for the UPS measurements, the freshly cleaved flakes were sputtered for ~2 minutes using an Ar source at 4 keV.

ASSOCIATED CONTENT

**Supporting Information**.

The Supporting Information is available free of charge via the Internet at http://pubs.acs.org.

TEM characterization of WTe$_2$ crystal; Atomic force microscopy (AFM) and Raman characterizations of exfoliated WTe$_2$ flakes; X-ray diffraction pattern (XRD) of exfoliated WTe$_2$ flakes; Transport data ($R_{xx}$(T), MR, and Hall) for two more thicknesses, 13 nm and 60 nm; Least square fit using the two-band model for Hall and MR curves in Figure 2; Temperature-dependent MRs and Hall curves, as well as the fitting result for the thick sample (115nm) and thin sample (13nm); Full UPS profiles of clean and 3.5 hours air-exposed WTe$_2$; 2D WAL for the ~ 2.5 nm sample.

AUTHOR INFORMATION

**Corresponding Author**

*E-mail: judy.cha@yale.edu.

**Present Addresses**

[+] (J. S.) University of Delft, Netherlands, J.Shen-1@tudelft.nl

**Author Contributions**

The manuscript was written through contributions of all authors. All authors have given approval to the final version of the manuscript. ‡J. M. W. and J. S. contributed equally.

**Acknowledgements**

This work was supported by Department of Energy, Basic Energy Science, Award No. DE-SC0014476. J. S. was partially supported by NSF DMR 1402600. J. M. W. was partially supported by NSF EFMA 1542815. P.K. was supported by DOE DE-SC0014476. Characterization facilities used in this work were supported by the Yale Institute for Nanoscience and Quantum Engineering (YINQE), MRSEC DMR 1119826, and Yale West Campus Materials Characterization Core. Raman characterization was conducted at the Yale Institute for the Preservation of Cultural Heritage.

**Figures**

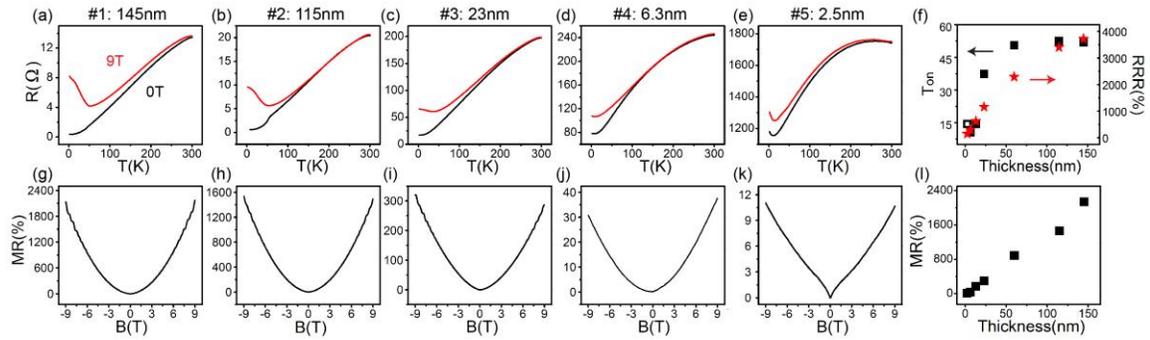

**Figure 1 | Thickness dependent transport properties of WTe$_2$ films: $T_{on}$, $RRR$ and MR.**
**(a)-(e)** R-T curves at 0 T and 9 T for samples with different thicknesses. **(f)** The $T_{on}$ and RRR extracted from these curves are plotted against thickness. We see that values for both $T_{on}$ and RRR systematically decrease for decreasing thickness. **(g)-(k)** the corresponding MR curves of the samples on the upper panel at T= 1.8 K. MR at 9 T as a function of thickness is summarized in (l), mirroring the effects seen on $T_{on}$ and RRR; the magnitude of the MR decreases dramatically as thinner WTe$_2$ flakes are measured. For the thinnest flake (device #5, 2.5 nm thickness) WAL is observed in the resistance vs. magnetic field plot.

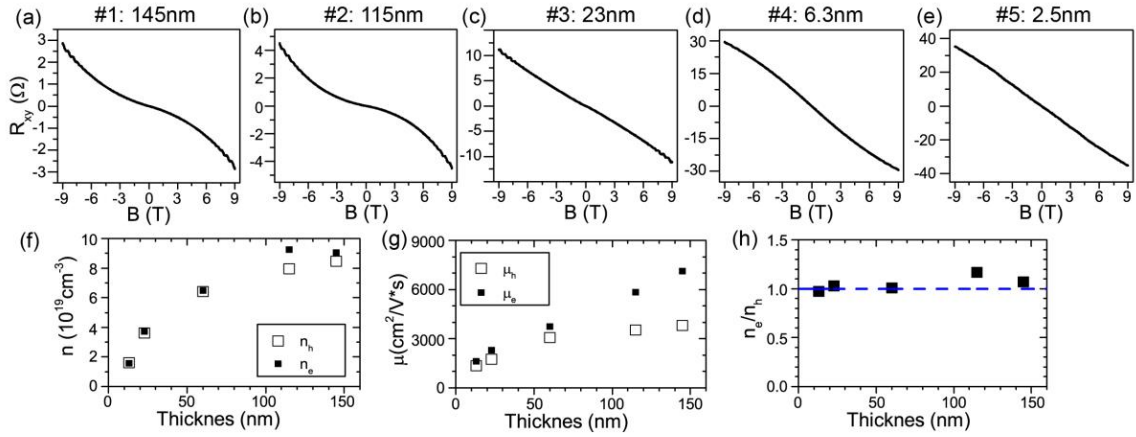

**Figure 2 | Thickness dependent mobility and carrier density from a two-band model.**
**(a)-(e)** Hall curves measured at 1.8 K for samples with different thicknesses. **(f)** The carrier densities and **(g)** mobilities of the electrons (solid squares) and holes (open squares) from the two band model fits of the Hall and MR curves summarized as a function of thickness down to 13 nm. Comparison of the data to the fit is shown in Figure S5. The fits for the 6.3 nm and 2.5 nm flakes are not shown due to the addition of WAL behavior in the $R_{xx}$ data. **(h)** Ratio of the electron carrier density, $n_e$, to the hole carrier density, $n_h$, as a function of thickness. Carrier compensation holds down to the 13 nm flake.

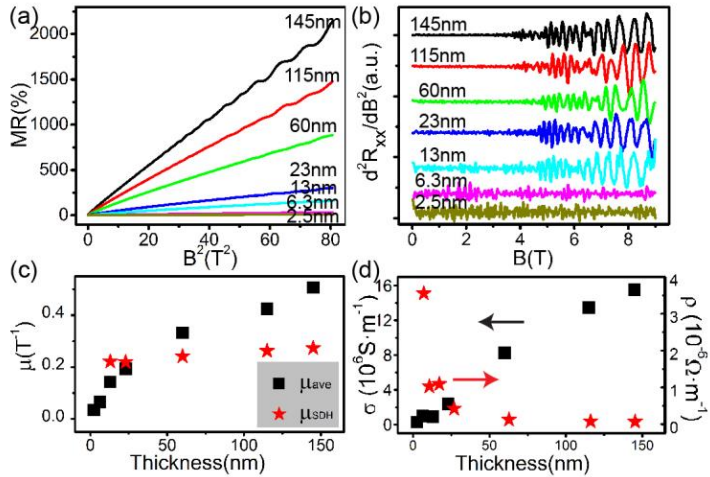

**Figure 3 | MR versus $B^2$ and SdH oscillations for WTe$_2$ flakes of different thicknesses.** **(a)** MRs measured at T= 1.8 K plotted as a function of $B^2$. The curves are linear as expected from the classical Lorentz law for MR. **(b)** SdH oscillations observed in samples thicker than 13nm disappear in thinner samples. **(c)** The average mobility calculated from MR ($\mu_{avg}$) and the mobility of the fastest carriers calculated based on the onset of the SdH oscillations ($\mu_{SdH}$) plotted together as a function of thickness. **(d)** The thickness dependence of the conductivity and resistivity of samples is consistent with that of $\mu_{avg}$.

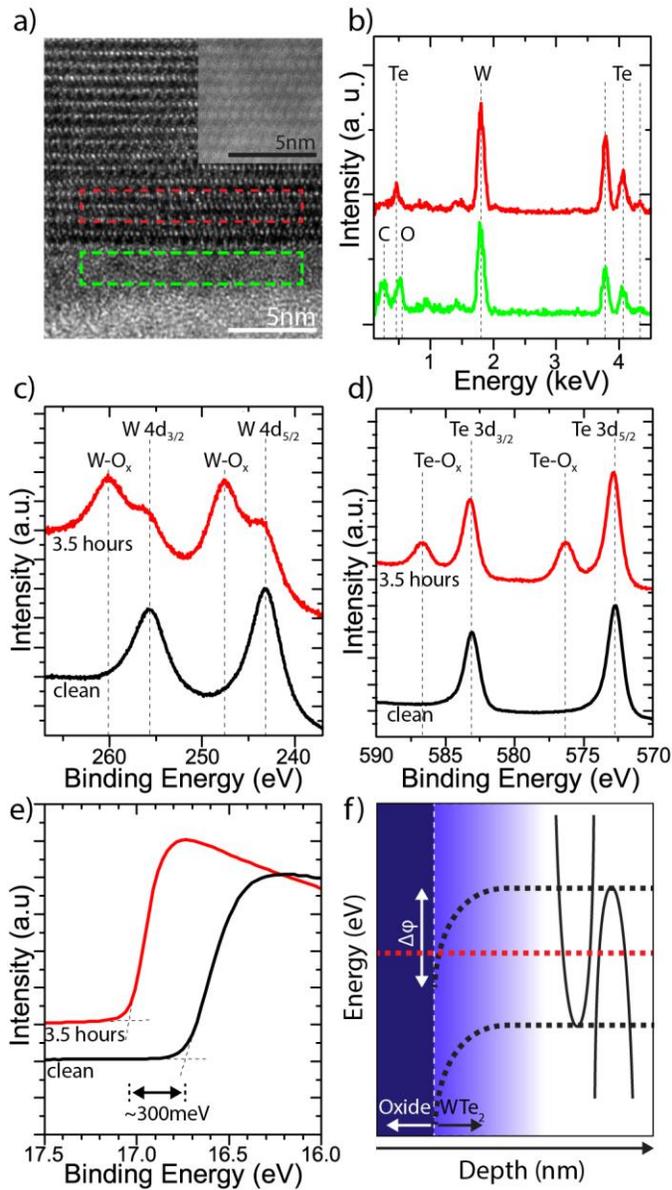

**Figure 4 | STEM characterization, X-ray photoelectron spectroscopy (XPS) and ultraviolet photoelectron spectroscopy (UPS).** (a) Cross-sectional TEM image of WTe$_2$. The inset shows the corresponding STEM annular dark field image. (b) EDX spectra of red and green dashed-line regions of (a). Oxygen is present only in the green region indicating this is an amorphous tungsten-tellurium oxide layer. This amorphous oxide layer is also present on the top surface of the flake. The bottom-most brighter amorphous layer is SiO$_2$

(c,d) XPS profiles for W (c) and Te (d) on clean flakes (black) and flakes with 3.5 hours of exposure to air (red). The presence of Te-$O_x$ and W-$O_x$ peaks after 3.5 hours of air exposure indicates formation of an oxide layer on the surface of the $WTe_2$ flakes. (e) UPS profiles of the secondary edge for clean (black) and 3.5 hours air exposed (red) $WTe_2$ flakes. The location of the secondary edge increases by about 300 meV, indicating a corresponding work function, $\varphi$, decrease of about 300 meV. (f) Schematic of Fermi level pinning between the surface oxide (left) and $WTe_2$ flake (right). The Fermi level is marked by a red dashed line. The Valence and conduction bands are shown on right. Black dashed lines track the location of the top/bottom of the valence/conduction bands in the proximity of the oxide interface.